\documentclass[aps,pra,twocolumn,groupedaddress,showpacs,showkeys]{revtex4}

\usepackage{graphicx}
\begin{document}

\newcommand{\be}{\begin{equation}}
\newcommand{\ee}{\end{equation}}

\title{A theory of electromagnetic fluctuations  for metallic surfaces and  van der Waals interactions between metallic bodies.}

\author{ Giuseppe Bimonte}
\email[Bimonte@na.infn.it]
\affiliation{Dipartimento di Scienze Fisiche, Universit\`{a} di
Napoli Federico II, Complesso Universitario MSA, Via Cintia
I-80126 Napoli, Italy;\\ INFN, Sezione di Napoli, Napoli, ITALY\\
}

\date{\today}

\begin{abstract}

A new general expression is derived for the fluctuating
electromagnetic field outside a metal surface, in terms of its
surface impedance. It  provides a generalization to real metals
of Lifshitz theory of molecular interactions between dielectric
solids.  The theory is used to  compute the radiative heat
transfer between two parallel metal surfaces at different
temperatures. It is shown that a measurement of this quantity  may
provide an experimental resolution of a long-standing controversy
about the effect of thermal corrections on the Casimir force
between  real metal plates.

\end{abstract}


\maketitle

One of the most intriguing predictions of Quantum Electrodynamics
is the existence of irreducible fluctuations of the
electromagnetic (e.m.) field  in the vacuum. It was Casimir's
fundamental discovery \cite{casimir} to realize that the effects
of this purely quantum phenomenon were not confined to the atomic
scale, as in the Lamb shift, but would rather manifest themselves
also at the macroscopic scale, in the form of a force of
attraction between two parallel discharged metal plates at
distance $L$.
By modern experimental techniques the Casimir force has now been
measured with an accuracy of a few percent (see Refs.\cite{decca}
and Refs. therein). For a recent review on both theoretical and
experimental aspects of the Casimir effect, see Ref.\cite{bordag}.
A few years after Casimir's seminal paper, E.M. Lifshitz developed
the general theory of molecular attractive forces between
dielectric bodies \cite{lifs} based on Rytov's theory of e.m.
fluctuations \cite{rytov}. In this theory, the physical origin of
the attraction resides in the fluctuating e.m.\ field which is
always present in the interior of any absorbing medium, and
extends beyond its boundaries partly in the form of propagating
waves and partly as evanescent waves. The resulting expression of
the attractive force   automatically includes both non-retarded
and retarded effects, as well as the effect of the bodies
temperature . Moreover, for zero temperature and in the limit
$\epsilon(\omega)= \infty$, it reproduces the force  found by
Casimir for two perfect mirrors.

The  basic assumption in Lifshitz theory is that one can describe
the propagation of e.m.\  fields inside the materials by means of
a dielectric constant. As it is well known \cite{london,landau},
this is not always the case for metals, because when the
penetration depth $\delta$ of the e.m.\  field becomes comparable
or larger than the free mean path of the conduction electrons,
there are non-local correlations in the material, and it is no
longer possible to use the simple form of the wavevector derived
from a dielectric response (anomalous skin-effect).  It is well
known, however that in a wide range of frequencies, including the
anomalous region, the fields {\it outside} a metal can be
accurately described by means of the {\it surface impedance}
$\zeta(\omega)$ introduced by M. A. Leontovich \cite{landau}. We
recall that the surface impedance $\zeta$ relates the tangential
components $E_t$ and $B_t$ of the e.m.\  fields on the surface of
the metal. For isotropic surfaces: \be {\bf E}_t=
\zeta(\omega)\,[{\bf B_t} \times {\bf n}]\;,\ee where ${\bf n}$ in
the  inward unit normal to the surface. For an ideal metal
$\zeta=0$, while for a good conductor $|\zeta| \ll 1$.

The purpose of this Letter is to develop the theory of e.m.\
fluctuations for metal surfaces. The main result is a new formula
for the correlation functions of the random e.m.\  fields that are
present {\it outside} a metal surface in thermal equilibrium, as a
result of the fluctuating microscopic currents in the interior of
the metal. A key feature is that the correlation functions are
expressed in terms of the surface impedance, and therefore they
apply to the anomalous region, as well as to superconductors
(extreme anomalous effect).

Let the metal occupy the $z <0$ half-space. We consider the
Fourier decomposition of the e.m.\  field outside the metal. For
the electric field of TE modes, we write: \be
\vec{E}^{(TE)}=\frac{1}{(2 \pi)^{3/2}}\int_0^{\infty} d \omega
\int d^2 k_{\perp}\,a(\omega,\vec{k}_{\perp})\,{\vec e}_{\perp}
e^{i( \vec{k}\cdot \vec{x}-\omega t)}+ c.c.\;, \ee where
$\vec{k}_{\perp}$ is the tangential component of the wave-vector
$\vec{k}$ and ${\vec e}_{\perp}$ is a unit vector perpendicular to
the plane formed by $\vec{k}_{\perp}$ and the normal  to the metal
surface. The third component of the wave-vector
$k_z=\sqrt{\omega^2/c^2-k_{\perp}^2}$ is defined such that
$Re(k_z)\ge 0$ and $Im(k_z) \ge 0$. In this way the external field
appears as a superposition of  {\it propagating waves} (p.w.)
travelling away from the surface (for $k_{\perp} < \omega/c$) and
of {\it evanescent waves} (e.w.) (for $k_{\perp} > \omega/c$)
exponentially dying out away from the surface. Similarly, we write
for the magnetic field of TM modes: \be \vec{B}^{(TM)}=\frac{1}{(2
\pi)^{3/2}}\int_0^{\infty} d \omega \int d^2
k_{\perp}\,b(\omega,\vec{k}_{\perp})\,{\vec e}_{\perp} e^{i(
\vec{k}\cdot \vec{x}-\omega t)}+ c.c.\;. \ee

Our new formulae for the correlation functions of the  fluctuating
e.m.\  field are expressed in terms of the statistical averages of
the products  of the amplitudes $a(\omega,\vec{k}_{\perp})$ and
$b(\omega,\vec{k}_{\perp})$:
\begin{widetext}\be \langle
a(\omega,\vec{k}_{\perp})\, a^*(\omega',\vec{k}'_{\perp}) \rangle=
 \frac{4 \pi \,\hbar\, \omega}{c}\, \coth \left(\frac{\hbar \omega}{2 k T}\right)\,\frac{Re(\zeta)} {|1+\zeta
\, k_z|^2}\,
\delta(\omega-\omega')\,\delta(\vec{k}_{\perp}-\vec{k}'_{\perp})\;,\label{corru}
\ee \be \langle b(\omega,\vec{k}_{\perp})\,
b^*(\omega',\vec{k}'_{\perp}) \rangle= \frac{4 \pi \,\hbar
\,\omega}{c}\,\coth \left(\frac{\hbar \omega}{2 k T}\right)\,
\frac{Re(\zeta)}{|\zeta+ \, k_z|^2}\,
\delta(\omega-\omega')\,\delta(\vec{k}_{\perp}-\vec{k}'_{\perp})\;,\ee
\be \langle a(\omega,\vec{k}_{\perp})\,
a(\omega',\vec{k}'_{\perp}) \rangle =\langle
b(\omega,\vec{k}_{\perp})\, b(\omega',\vec{k}'_{\perp})
\rangle=\langle a(\omega,\vec{k}_{\perp})\,
b^*(\omega',\vec{k}'_{\perp}) \rangle= \langle
a(\omega,\vec{k}_{\perp})\, b(\omega',\vec{k}'_{\perp}) \rangle
=0\;, \label{corrt}\ee \end{widetext}   with $k$ Boltzmann's
constant. We obtained the above formulae by considering p.w. with
frequencies $\omega$ belonging to the  domain of the normal skin
effect, for which dielectric  and surface impedance b.c. give
coinciding results provided only that $ \epsilon'' \gg 1$, with
$\epsilon''$ the imaginary part of $\epsilon$. In this region
\cite{landau} one can take $\zeta=\sqrt{\mu/\epsilon}$, with $\mu$
the magnetic permeability (we took $\mu=1$) and  then, for p.w. in
this domain, Eqs.(\ref{corru}-\ref{corrt}) coincide to high
accuracy with the analogous expressions derived from Lifshitz
theory. We stress however that the validity of Eqs.
(\ref{corru}-\ref{corrt}) extends to the anomalous region, where
Lifshitz theory is inapplicable, and therefore they constitute  a
new non-trivial generalization of that theory. Details will be
given elsewhere.

As a first application of Eqs.(\ref{corru}-\ref{corrt}), we have
obtained a new derivation  of the Casimir force between two
metallic plates. Existing derivations based on surface impedance
b.c., as found in the literature (see for example
Ref.\cite{bezerra}), use a mode analysis  to evaluate the Casimir
free energy of the cavity, much in the spirit of the original
Casimir's paper. The  delicate thing in this approach is to
justify the validity of the result in the case of dissipative
mirrors, which requires consideration of an auxiliary
electrodynamic problem (see the second of Ref. \cite{bezerra} for
details). Using Eqs. (\ref{corru}-\ref{corrt}), we can however
obtain the Casimir force in much the same way used by Lifshitz,
i.e. by computing the statistical average of the $z$-component of
the Maxwell stress tensor inside the gap, resulting from
fluctuating e.m. fields, at a point close to the surface of either
plate. Dissipation poses no problem now, and is taken into account
from the start. The computation is rather lengthy and gives the
following result for the spectral density $F_{\omega}$ of the
attraction force $F$ per unit area ($F=\int_0^{\infty}d\omega\,
F_{\omega}$): \be F_{\omega}= \coth \left(\frac{\hbar \omega}{2 k
T}\right)\,\frac{\hbar \omega^3}{2\pi^2 c^3} Re \int p^2 dp
\sum_{\alpha }[(C_{\alpha}-1)^{-1}+1/2]\;,\label{vdw}\ee where \be
C_{\alpha}=e^{-2 i p\, \omega\, L/c}/(r^{(1)}_{\alpha}\,
r^{(2)}_{\alpha} )\;\;,\;\;\;\alpha=TE,\,TM.\label{ci} \ee Here
$p=c\, k_z/{\omega}$ and the integration with respect to $p$ is
along the real axis from 1 to 0 (corresponding to p.w.) and thence
along the imaginary axis to $i \infty$ (corresponding to e.w.),
while $r^{(1)}_{\alpha}$ and $r^{(2)}_{\alpha}$ are the impedance
reflection coefficients for mirror 1 and 2 respectively: \be
r_{TM}^{(i)}=\frac{c \,k_z/\omega -\zeta_i}{c \,k_z/\omega
+\zeta_i} \;\;,\;\;\;\;r_{TE}^{(i)}=\frac{  \zeta_i \,c
\,k_z/\omega-1 }{\zeta_i \,c \,k_z/\omega+1}\;\label{refc}. \ee
Eqs. (\ref{vdw}) and (\ref{ci}) are of the same form as those
obtained by Lifshitz, apart from the expression of the reflection
coefficients. After disregarding in Eq.(\ref{vdw}) the terms with
$1/2$, which give a divergent $L$-independent contribution, and
after a rotation of the integration contours of $\omega$ and $p$,
our expression for $F$ becomes identical to that in
\cite{bezerra}.

The second application of Eqs.(\ref{corru}-\ref{corrt}) that we
consider is new and much more interesting, as it relates to the
presently controversial issue of the modification of the Casimir
force arising from a non zero temperature of the mirrors, when the
latter are treated as real metals. The debate was raised by the
findings of Refs.\ \cite{sernelius}, showing that the combined
effect of temperature and finite conductivity leads to large
deviations from the ideal metal case. This result was obtained
within the framework of Lifshitz theory, by using the Drude
dielectric function $\epsilon_D(\omega$)
\be\epsilon_D(\omega)=1-{\Omega_p^2}/[\,\omega \,(\omega + i
\gamma)]\;\label{drude}\ee  to describe the metal. Finite
conductivity is taken into account by allowing a non-vanishing
value for the relaxation frequency $\gamma$.  The results of
Refs.\ \cite{sernelius} have been criticized by several authors,
and supported by others. There is no space here to discuss the
matter, and we address the reader to Refs. \cite{bezerra} and
\cite{hoye} for a discussion of the extended literature on this
topic.
\begin{figure}
\includegraphics{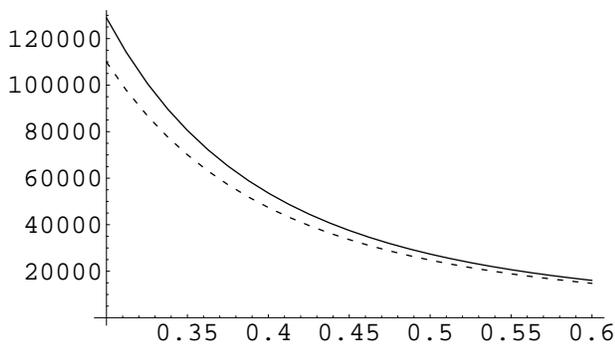}
\caption{\label{plotssigma}  Plots of radiative heat transfer
between two Al plates (in erg/(s cm$^2$)) at temperatures
$T_1=323$ K and $T_2=300$ K, as a function of separation (in
microns), according to Lifshitz theory (dashed line) and surface
impedance approach (continuous line).}
\end{figure}

Of particular  interest to us is the criticism of
Ref.\cite{sernelius} in the second of Refs.\cite{bezerra}, where
it is argued that the dielectric model, {\it in combination with
the Drude dielectric function}, fails to describe the thermal
Casimir force {\it even in the domain of the normal skin effect},
and that also in this region  the appropriate description is
provided by impedance b.c.. Further arguments in favor of this
point of view were found  in \cite{lamor},  where the spectrum of
finite temperature corrections to the Casimir force is studied,
for metallic plates of finite conductivity.   This study shows
that the large deviations from the ideal metal case in
\cite{sernelius} arise from {\it thermal TE e.w.\ of low
frequencies} ($\omega=10^{10} - 10^{13}$ rad/s for $L=1 \mu$m), for which the description of the plates as bulk dielectrics is not valid. It is also shown   there that
if   surface impedance b.c. are used, with   $\zeta=
1/\sqrt{\epsilon_D}$ (which is the expression valid in the domain
of the normal skin effect), instead of the large {\it repulsive}
thermal correction from TE e.w.\ found in \cite{sernelius},  one
obtains an {\it attractive} correction, of much smaller magnitude,
while no appreciable differences are found both  in the TM  and in
the TE p.w. sectors \footnote{See Ref. \cite{bimonte}, where a
mistake in impedance b.c. for TE modes made in \cite{lamor} is
corrected.}. The conclusion of this study is that the conflicting
statements on thermal corrections to the Casimir force, may be
traced back to different magnitudes  for the thermal TE e.w.
correction, in the two approaches.

In view of these findings, it would be important to devise new
experiments to probe the effectiveness of the two approaches to
describe physical effects  of thermal TE e.w., in the frequency
range that is relevant for the Casimir effect. Optical data are of
no help here, since they refer to p.w., and therefore give no
information on e.w.. A key remark now is that the relevant e.w.
are not vacuum fluctuations, as in the Casimir force at zero
temperature, but rather real thermal excitations. Now, it is known
that thermal e.w. play an important role in the heat transfer $S$
between two closely spaced metal surfaces, at temperatures $T_1$
and $T_2$ \cite{polder}. It is therefore very interesting  to see
what is the prediction of impedance theory for $S$, and to compare
it with the result from Lifshitz theory.  The computation of $S$
goes through the same steps  as in Lifshitz' computation of the
Casimir force \cite{polder}. One assumes that each plate
separately is in local thermal equilibrium, and emits a
fluctuating e.m. field according to our Eqs.
(\ref{corru}-\ref{corrt}), with $T$ the temperature of the
emitting plate. Since  thermal e.m. fields from plates 1 and 2 are
uncorrelated, the radiated power $S$  can be obtained by
subtracting incoherently the average Poynting vectors for the
multiply reflected radiation fields originating from either plate.
Apart from the fact that we obviously evaluate the average
Poynting vector instead of the Maxwell stress-tensor, the only
difference with respect to the computation of thermal corrections
to the Casimir force is that the fluctuating fields from plate 1
and 2 are relative now to distinct temperatures $T_1$ and $T_2$.
Apart from this, these thermal fields have the same expression as
in the Casimir force computation, at nonzero temperature. Our
final formula for the power $S$ (per unit area) has the form of a
difference between two terms, one for each plate:
\begin{widetext} \be S=-\frac{4 \hbar}{\pi^2 c^2} \int_0^{\infty}
d \omega \,\omega^3\left(\frac{1}{\exp (\hbar \omega/ k
T_1)-1}-\frac{1}{\exp (\hbar \omega/ k T_2)-1}\right){{\rm
Re}(\zeta_1) {\rm Re}(\zeta_2) }\,{\rm Re}\,\int dp \,p\, |p|^2
\,|e^{2 i p L \omega/c }|\left(\frac{1}{B_{TE}}+
\frac{1}{B_{TM}}\right)\, \label{htim}\ee
\end{widetext}
where the contour of integration for $p$ is the same as in
Eq.(\ref{vdw}), and the quantities $B_{TE/TM}$ are defined as: \be
B_{TE}=|(1+p \zeta_1)(1+p \zeta_2)-(1-p \zeta_1)(1-p \zeta_2) \exp
(2 i p L \omega/c)|^2\;, \ee \be B_{TM}=|(p+ \zeta_1)(p+
\zeta_2)-(p- \zeta_1)(p- \zeta_2) \exp (2 i p L \omega/c)|^2\;.
\ee  In Fig.\ 1 we show plots of the radiated power (in erg/(s
cm$^2$)), for two plates of Al as a function of the separation $L$
(in microns), for $T_1=323$ K and $T_2=300$ K. In parallel with
the analysis of thermal corrections to the Casimir force described
earlier,  the figure compares the result from Lifshitz theory, Eq.
(19) of Ref. \cite{polder}, with the Drude dielectric function Eq.
(\ref{drude}) (dashed line), with our Eq.\ (\ref{htim}), with
$\zeta(\omega)=1/\sqrt{ \epsilon_D (\omega)}$ (solid line). In
both cases, we took $\hbar \,\Omega_p=11.5$ eV and $\hbar
\gamma=.05$ eV. We see that impedance theory and Lifshitz theory
give rather different predictions for  $S$ (for $L=0.3\,\mu$m,
impedance theory gives for $S$ a value approximately  17 \% larger
than Lifshitz theory ). We observe that in the considered
submicron separation range, heat transfer occurs mainly via TE
e.w..  The smaller contribution from p.w. and from TM e.w.\ is
practically the same for the two theories, to an accuracy of a few
parts in ten thousand, and therefore the difference between the
two curves in Fig.\ 1 arises exclusively from TE e.w..
 For larger separations the relative weight of e.w.\ with
respect to p.w.\ decreases, and the two theories lead to very
similar results. Already at one micron the results from Lifshitz
theory and impedance theory differ only by four percent.
\begin{figure}
\includegraphics{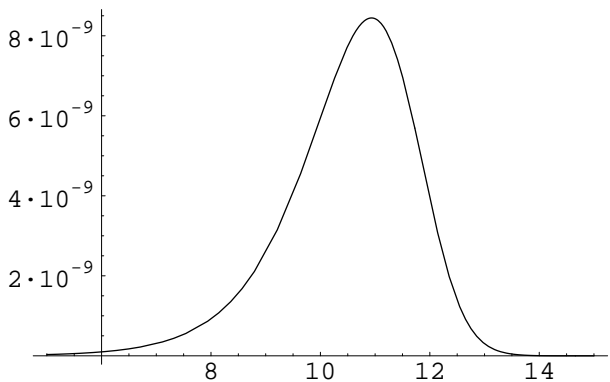}
\caption{\label{plotssigma} Plot of  the difference $s^{(\rm
imp)}_{TE \, e.w.}(\omega) - s^{(\rm Lif)}_{TE \, e.w.}(\omega)$
between TE e.w.\ contributions to the spectra $s^{\rm
(imp)}(\omega)$ (impedance theory) and $s^{\rm (Lif)}(\omega)$
(Lifshitz theory) of radiative heat transfer   between two Al
plates (in erg/(rad cm$^2$)) at temperatures $T_1=323$ K and
$T_2=300$ K, as a function of $\log_{10}(\omega)$. $L=0.3$
$\mu$m.}
\end{figure}
We evaluated   the contribution $s_{TE\,e.w.}(\omega)$ from TE
e.w.\ to the spectra $s(\omega)$ of the radiated power
($S=\int_0^{\infty} d \omega \, s(\omega)$) for the two theories
and in Fig. 2 we plot the difference $ s^{(\rm
imp)}_{TE\,e.w.}(\omega) - s^{(\rm Lif)}_{TE\,e.w.}(\omega)$, as a
function of $\log_{10}(\omega)$, for $L=0.\,3\, \mu$m. We see that
two spectra differ significantly precisely in the low-frequency
region which, according to \cite{lamor} is at the origin of the
controversial thermal corrections to the Casimir force.
As pointed out in Ref.\cite{lamor}, in this frequency domain
the penetration depth is of the same order as the mean free path of the conduction
electrons. Therefore the
dielectric model for the plates is not valid, and a more relistic
description is provided by surface impedance b.c.. It is  clear that an
accurate
measurement of $S$ would provide strong indications of which approach
is preferable, a knowledge which could then be used in the evaluation of
the thermal Casimir effect.
An experiment
measuring the radiative heat transfer between two chromium plates
is described in Ref.\ \cite{harg}. Even though the
author of Ref. \cite{harg} claims to have obtained some evidence
for proximity effects related to e.w., his measurements were
performed mostly for plates separations larger than
one micron, where we expect small differences between the two
theories. Moreover, as the author does not quote the area of the
plates, no comparison is possible with his data. It would
therefore be desirable to repeat the measurements for submicron
separations, and for metals that are used in Casimir experiments.

In conclusion, we have presented  new formulae for the correlation
functions of e.m.\ fluctuations present outside a metal. The
formulae involve the surface impedance of the metal, and are
therefore applicable in the anomalous region, as well as in the
extreme anomalous case (superconductors), where Lifshitz theory is
not valid. As an application, we have evaluated the radiative heat
transfer between two metal plates at different temperatures and we
have shown that a measurement of this quantity should provide
enough information to settle experimentally recent controversies
about the thermal Casimir effect. We plan to apply our results to
superconducting cavities, that were proposed recently \cite{bimo}
as a tool to measure  the variation of Casimir energy across the
superconducting phase transition.

The author thanks G.L. Klimchitskaya and V.M. Mostepanenko for
valuable suggestions, and acknowledges partial financial support
by PRIN {\it SINTESI}.


\begin{thebibliography}{200}

\bibitem{casimir}  H.B.G.~Casimir, Proc. K. Ned. Akad. Wet.
Rev. {\bf 51}, 793 (1948).



\bibitem{decca}S.K. Lamoreaux, Phys. Rev. Lett. {\bf 78}, 5 (1997);  U. Mohideen and A. Roy, ibid. {\bf 81},
4549 (1998); G. Bressi, G. Carugno, R. Onofrio and G. Ruoso, ibid.
{\bf 88} 041804 (2002); R.S. Decca, D. L\'opez, E. Fischbach, and
D.E. Krause, ibid. {\bf 91}, 050402 (2003).

\bibitem{bordag} M. Bordag, U. Mohideen and V.M. Mostepanenko,
Phys. Rep. {\bf 353}, 1 (2001).

\bibitem{lifs} E.M. Lifshitz, Sov. Phys. JETP {\bf 2}, 73 (1956);
E.M. Lifshitz and L.P. Pitaevskii, {\it Landau and Lifshitz Course
of Theoretical Physics: Statistical Physics Part II}
(Butterworth-Heinemann, 1980.)

\bibitem{rytov} S.M.\ Rytov, {\it Theory of Electrical Fluctuations and
Thermal Radiation}, Publyshing House, Academy os Sciences, USSR (1953).

\bibitem{london} H. London, Proc. R. Soc. {\bf A 176}, 522 (1940).

\bibitem{landau}
L.D. Landau and  E.M. Lifshitz, {\it Landau and Lifshitz Course
of Theoretical Physics: Electrodynamics of Continous
Media} (Oxford: Pergamon, 1960.)



\bibitem{bezerra}
V.B. Bezerra, G.L. Klimchitskaya and C. Romero, Phys. Rev. {\bf A
65} 012111 (2001); B. Geyer, G.L. Klimchitskaya and V.M.
Mostepanenko, ibid. {\bf A 67} 062102 (2003); V.B. Svetovoy and
M.V. Lokhanin, Phys. Rev. {\bf A 67}, 022113 (2003).

\bibitem{sernelius} M. Bostrom and B.E. Sernelius,
Phys. Rev. Lett. {\bf 84}, 4757 (2000); B.E. Sernelius, {\it
ibid.} {\bf 87}, 139102 (2001).

\bibitem{hoye}   J.S. H$\o$ye,
I. Brevik, J.B. Aarseth and K.A. Milton, Phys. Rev. {\bf E 67},
056116 (2003) and references therein.

\bibitem{lamor} J.R. Torgerson and S.K. Lamoreaux,  Phys. Rev. {\bf E 70},
 047102 (2004); S.K. Lamoreaux, Rep. Prog. Phys. {\bf 68}, 201
(2005).

\bibitem{bimonte} G. Bimonte, {\it Comment on "Low-frequency character of the Casimir force between metallic
films"}, in press on Phys. Rev. {\bf E}, (2006).

\bibitem{polder} D. Polder and M. Van Hove, Phys. Rev. {\bf B 4},
3303 (1971).

\bibitem{harg} C.M. Hargreaves, Phys. Lett {\bf A 30}, 491 (1969).



\bibitem{bimo} G.Bimonte, E. Calloni, G. Esposito, L. Milano and L.
Rosa, Phys. Rev. Lett. {\bf 94}, 180402 (2005); G. Bimonte, E.
Calloni, G. Esposito, and L. Rosa, Nucl. Phys. {\bf B 726}, 441
(2005).






\end{thebibliography}
\end{document}